\documentclass[epjST]{svjour}
\usepackage{color}
\usepackage{graphics}
\usepackage{graphicx,epstopdf}
\usepackage{latexsym}
\begin{document}
\title{Enhancing synchronization in chaotic oscillators by induced heterogeneity}
\author{Ranjib Banerjee\inst{1}, Bidesh K. Bera\inst{2}, Dibakar Ghosh\inst{2}\fnmsep\thanks{\email{diba.ghosh@gmail.com}} \and Syamal Kumar Dana\inst{3}}
\institute{\inst{1}School of Engineering \& Technology, BML Munjal University, Gurgaon, India \\
	\inst{2}Physics and Applied Mathematics Unit, Indian Statistical Institute, Kolkata, India\\\inst{3}Department of Mathematics, Jadavpur University, Kolkata 700032, India}
%

\abstract{
 We report enhancing of complete synchronization in identical chaotic oscillators when their interaction is mediated by a mismatched oscillator. The identical oscillators now interact indirectly through the intermediate relay oscillator.  The induced heterogeneity in the intermediate oscillator plays a constructive role in reducing the critical coupling for a transition to complete synchronization. A common lag synchronization  emerges between the mismatched relay oscillator and its neighboring identical oscillators that leads to this enhancing effect. 
We present examples  of one-dimensional open array, a ring, a star network and a two-dimensional  lattice of dynamical systems to demonstrate how this enhancing effect occurs. The  paradigmatic  R\"ossler oscillator is used as a dynamical unit, in our numerical experiment, for different networks to  reveal the enhancing phenomenon.  }

\maketitle

\section{Introduction}
\label{intro}
 Various types of collective behaviors emerge when two or more oscillatory units  interact with each other; synchronization is one of the most important collective behaviors due to a range of applications in different fields \cite{bocaletti,application}, physics, chemistry, biology, network science and technology. The concept of chaos synchronization  is useful in secure communications \cite{communi1,communi2,communi_pre,communi_epl1,communi_epl2}, for encoding and information processing in neuronal systems. Synchronization  in two coupled chaotic oscillators and  also in complex networks has been extensively studied in the last two decades \cite{kurths,phys_report}. Some of the important forms of synchronization are complete synchronization (CS) or zero lag synchronization (ZLS) \cite{relay1}, phase synchronization  (PS) \cite{phase_prl}, lag synchronization (LS) \cite{lag_prl} and generalized synchronization (GS) \cite{genl_pre} that might occur either in identical or nonidentical chaotic systems. In a large ensemble of oscillators, more varieties of collective behaviors such as clustering \cite{cluster}, partial synchronization \cite{partial}, chimera states \cite{chimera,chimera2,chimera3},  relay synchronization (RLS) \cite{relay1,relay2,relay3,relay4} and remote synchronization \cite{remote1,remote2}, were reported. In this paper, we emphasize on RLS \cite{relay1} that defines a state of synchrony between two indirectly coupled oscillators interacting through an  intermediate oscillator, called as a relay unit in a network. We revisit our previous study \cite{relay2} on RLS and the related enhancing effect in an open array of oscillators and further extend the results  to different other networks. RLS was first reported \cite{relay1} in diode lasers when two delay coupled laser sources  were interacting via a relaying  third laser source. The robustness of RLS against  heterogeneity and noise were experimentally demonstrated in lasers \cite{relay1,laser1,laser2,laser3} and electronic circuits \cite{electronic1,electronic2}. The idea of RLS is useful \cite{communi3} for transmitting and recovering encrypted massages. The dynamical relaying mechanism and the associated RLS is a possible recipe for isochronous synchronization between distantly located cortical areas of brain \cite{brain1,brain2,brain3}.
\par 
We established a type of RLS earlier \cite{relay2} where  the critical coupling for CS was found reduced in two identical chaotic oscillators when the parameter of the  mediating third oscillator was detuned from the identical parametric condition. The induced parameter mismatch or heterogeneity in the mediating oscillator played a constructive role on CS of the indirectly interacting identical oscillators. Basically, LS  emerged between  both the identical units and the mismatched relay unit for a coupling strength lower than the coupling threshold for a CS state.  This reduction in critical coupling of CS between the identical units via the relay unit was explained as {\textit enhancing of synchrony}.  
We show here that the enhancing phenomenon is not limited to an open chain of oscillators rather it can emerge in other type of networks as well. RLS was also reported by others \cite{relay4} during the onset of GS  where the authors also found an enhancing effect. The role of heterogeneity in the enhancement of CS was also explored \cite{enh_boca} in a complex network. Different processes such as induced heterogeneity \cite{relay3,enh_boca}, coupling delay \cite{enh_delay}, noise \cite{diba_physica,diba,mixed_syn,senthil} have been suggested for the enhancement of chaos synchronization in two or more oscillators.
\par  We emphasize here that a natural presence of heterogeneity in dynamical systems or an induced heterogeneity (positive and negative) can really play constructive role on synchrony. We explore this constructive role in chaotic oscillators using different coupling configurations,
 an one-dimensional open array, a star network, a ring of oscillators and a 2-dimensional (2D) lattice of oscillators. 
We consider the chaotic R\"ossler oscillator as a paradigmatic model, althrough the text, and always use mutually diffusive  interactions between any two oscillators, although we know that the phenomenon can be seen in many other chaotic models. We mention here that the enhancing effect is observed in chaotic systems only \cite{relay2} since LS emerges in such systems at a lower coupling than the critical coupling for CS. On the other hand, in limit cycle systems, CS emerges  in identical systems  for a coupling smaller than the coupling for LS in mismatched systems. 
\par The rest of the paper is arranged as follows: the enhancement of CS under RLS  in one dimensional linear array, ring network are presented in section 2. The study has been extended  to star networks as illustrated in section 3. The case of two dimensional array  is explained in section 4.  The manuscript ends with a conclusion in section 5.

\section{One Dimensional  Array and a Ring Lattice}
\label{sec:1}
We consider an open array and a ring of N-coupled R\"ossler oscillators with nearest neighbor diffusive coupling. The dynamical equation of the  networks is
$$\dot x_i=-\omega_iy_i-z_i+\epsilon(x_{i-1}+x_{i+1}-2x_i)\;\;\;\;\;\;\;\;\;\;\;\;\;\;\;\;\;\;\;\;\;\;\;\;\;\;\;\;\;\;\;\;\;\;\;\;\;\;\;$$
$$ \dot y_i=\omega_i x_i+ay_i \;\;\;\;\;\;\;\;\;\;\;\;\;\;\;\;\;\;\;\;\;\;\;\;\;\;\;\;\;\;\;\;\;\;\;\;\;\;\;\;\;\;\;\;\;\;\;\;\;\;\;\;\;\;\;\;\;\;\;\;\;\;\;\;\;\;\;\;\;\;\;\;\;\;\;\; \eqno{(1)}$$
$$ \dot z_i =b+z_i(x_i-c),\;\;\;\;i=1,...,N.\;\;\;\;\;\;\;\;\;\;\;\;\;\;\;\;\;\;\;\;\;\;\;\;\;\;\;\;\;\;\;\;\;\;\;\;\;\;\;\;\;\;\;\;\;\;\;$$
where $\epsilon$ is the coupling strength. 
\begin{figure}
	\resizebox{0.95\columnwidth}{!}
	{\includegraphics{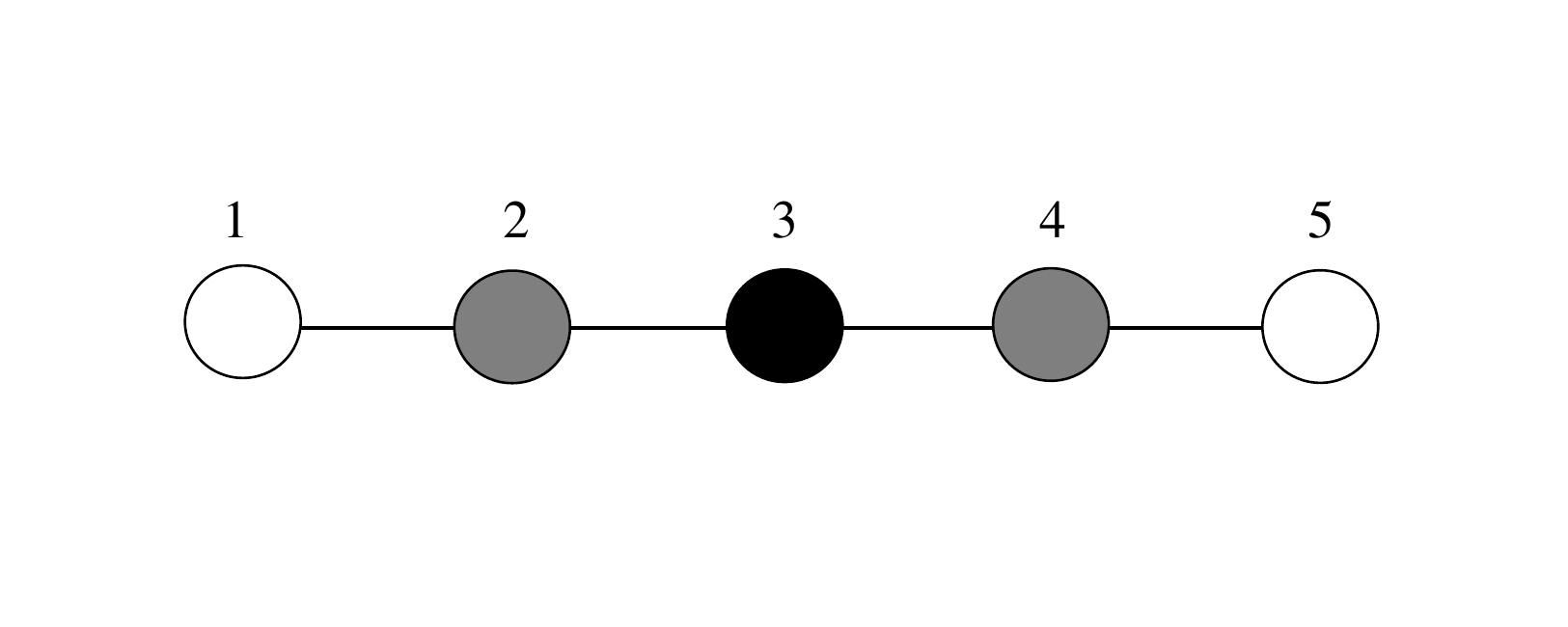} }
	\resizebox{1.1565\columnwidth}{!}
	{\includegraphics{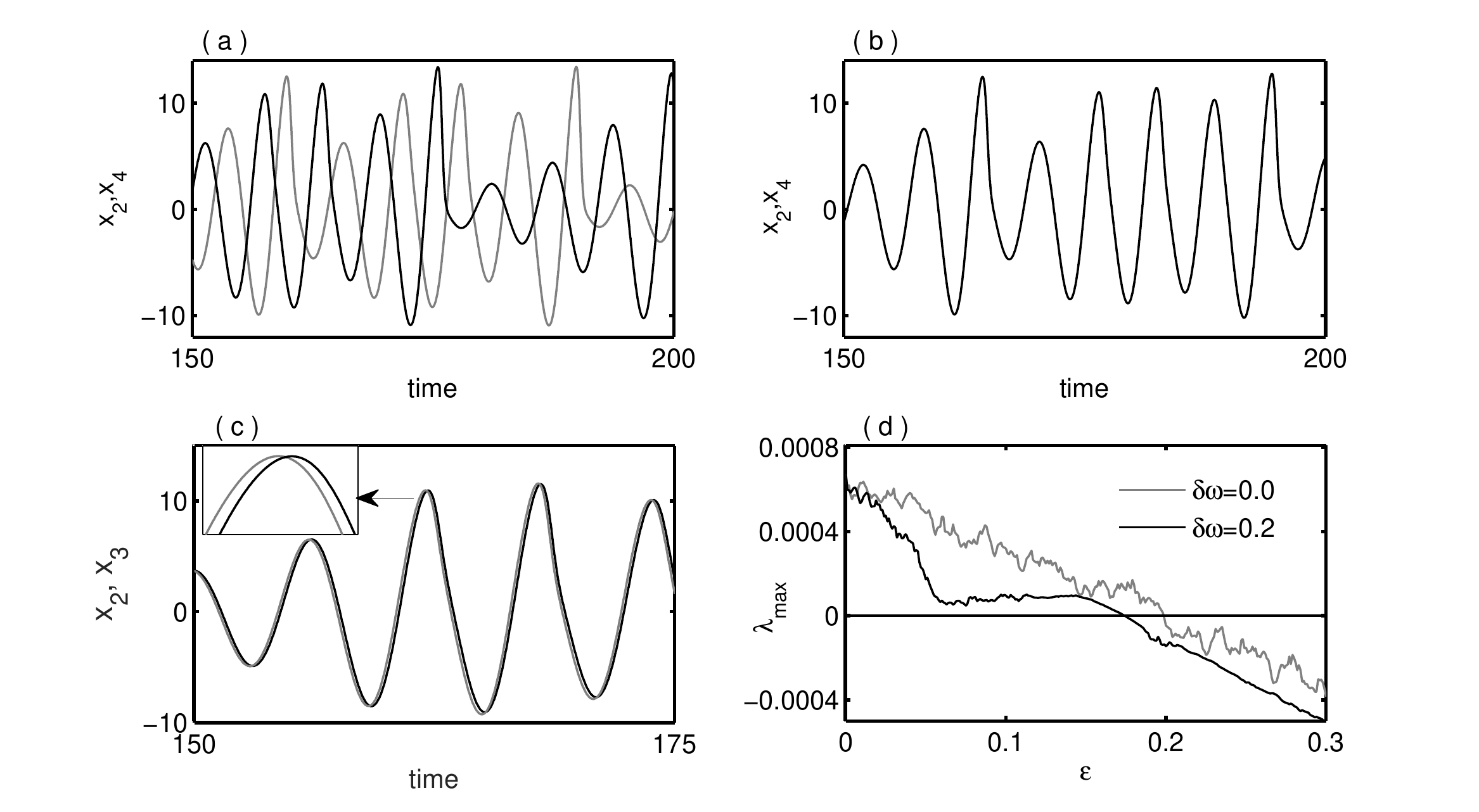} }
	\caption{ One-dimensional array of five R\"ossler oscillators (top). The oscillator in black color has parameter mismatch. Pair of $(x_2, x_4)$ times series shown for $\epsilon=0.19$, (a) in black and gray lines in all identical case when they are in a desynchronized state, (b) CS emerges when a mismatch $\delta \omega=0.2$ is introduced in the relay oscillator and, (c) pair of $(x_2, x_3)$ time series shows a LS state; time lag of the adjacent oscillators with the mismatched oscillator. (d) Plots of master stability function $\lambda_{max}$ against the coupling strength $\epsilon$ shows (gray line for $\delta \omega$=0, black line for $\delta \omega=0.2$) a drifting to a lower critical value for induced heterogeneity.}
	\label{timelag}
\end{figure}
 The uncoupled system ($\epsilon=0$) exhibits chaotic dynamics for a choice of parameters,  $\omega_i=\omega+\delta\omega_i$, $a=0.2$, $b=0.4$, and $c=7.5$.  For all the oscillators, $\omega=1$ and $\delta\omega=\delta\omega_i=0$, except for the mismatched oscillator, for which $\delta\omega=0.2$. For a negative mismatch, $\omega=1.1$ and $\delta\omega=-0.1$. Depending on the boundary conditions, Eqn.(1) represents  either an open array or a ring topology. We consider five ($N=5$) R\"ossler oscillators, as an example, where a boundary condition $x_0=x_1$ and $x_{N+1}=x_N$ represents an open array. For the ring coupling configuration, a periodic boundary condition  $x_0=x_N$ and $x_{N+1}=x_1$ is considered. An open array of five oscillators  is shown in the upper panel of Fig.~\ref{timelag} where the black circle represents the  relay oscillator with a mismatch $\delta\omega$ and indirectly coupled identical oscillators in open and gray circles. In this open array, the critical coupling for CS between the identical oscillators in symmetric positions to the relay oscillator in the center, reduces. A LS state emerges between the central relay oscillator and its neighboring relayed systems, and thereby CS is enhanced between all the identical oscillators in symmetric positions on both sides. All identical oscillators (N=5) emerge into a CS state for a coupling strength $\epsilon \ge 0.1985.$ Figure~\ref{timelag}(a) shows a desynchronized state between the indirectly coupled oscillators (2, 4) for a coupling strength $\epsilon=0.19$ that is lower than the critical coupling when all the oscillators are identical.  When a mismatch $\delta \omega=0.2$ is introduced in the central oscillator (black circle oscillator 3), then CS is observed between $(x_2, x_4)$- and $(x_1, x_5)$-pair and time lag is observed between the time series of $(x_2, x_3)$- or $(x_3, x_4)$-pair in Figs.~\ref{timelag}(b) and (c), respectively. The inset figure of Fig.~\ref{timelag}(c) shows the lag synchronization in zooming portion of the time series. Interesting point to note that the amount of delay does not propagate as the distance of the relayed systems increases from the relay unit. Rather, it remains constant for all the oscillators in the network. Figure~\ref{timelag}(d) shows the variation of master stability function (MSF) $\lambda_{max}$ with coupling strength $\epsilon$. Pairs of (2, 4)-oscillators and (1, 5)-oscillators  emerge into a CS state for $\epsilon>0.1985$ (gray line) when five oscillators are identical ($\delta \omega=0$). $\lambda_{max}$ crosses the line to a negative value at a lower critical coupling strength $\epsilon \ge 0.174$ (black line) when a mismatch is introduced in the central oscillator.  For a negative mismatch ($\delta \omega=-0.1$), the  effect is found similar as shown in Fig~\ref{neg-mis}.

\par Next we consider  a ring of oscillators shown in Fig.~\ref{ring}. The dynamical equation for this network is represented by Eqn.(1) with the boundary conditions, $x_0=x_N$ and $x_{N+1}=x_1$. For this ring network, five ($N=5$) mutually coupled R\"ossler oscillators are chosen, with any one of them having a parameter mismatch: here the black color circle is the mismatched oscillator with an amount of mismatch $\delta\omega$. The nodes at equal distances on both sides of the relay oscillator show an enhancing effect and the distance from the central node does not affect the time lag. Variation of $\lambda_{max}$ with respect to coupling strength $\epsilon$ is shown in the right panel of Fig.~\ref{ring} for identical (gray) and mismatched (black) cases. We observe again that the critical coupling strength for CS in all the identical oscillators is lowered when a mismatch is introduced in any one of the oscillator. 
\begin{figure}
	\resizebox{0.6\columnwidth}{!}
	{\includegraphics{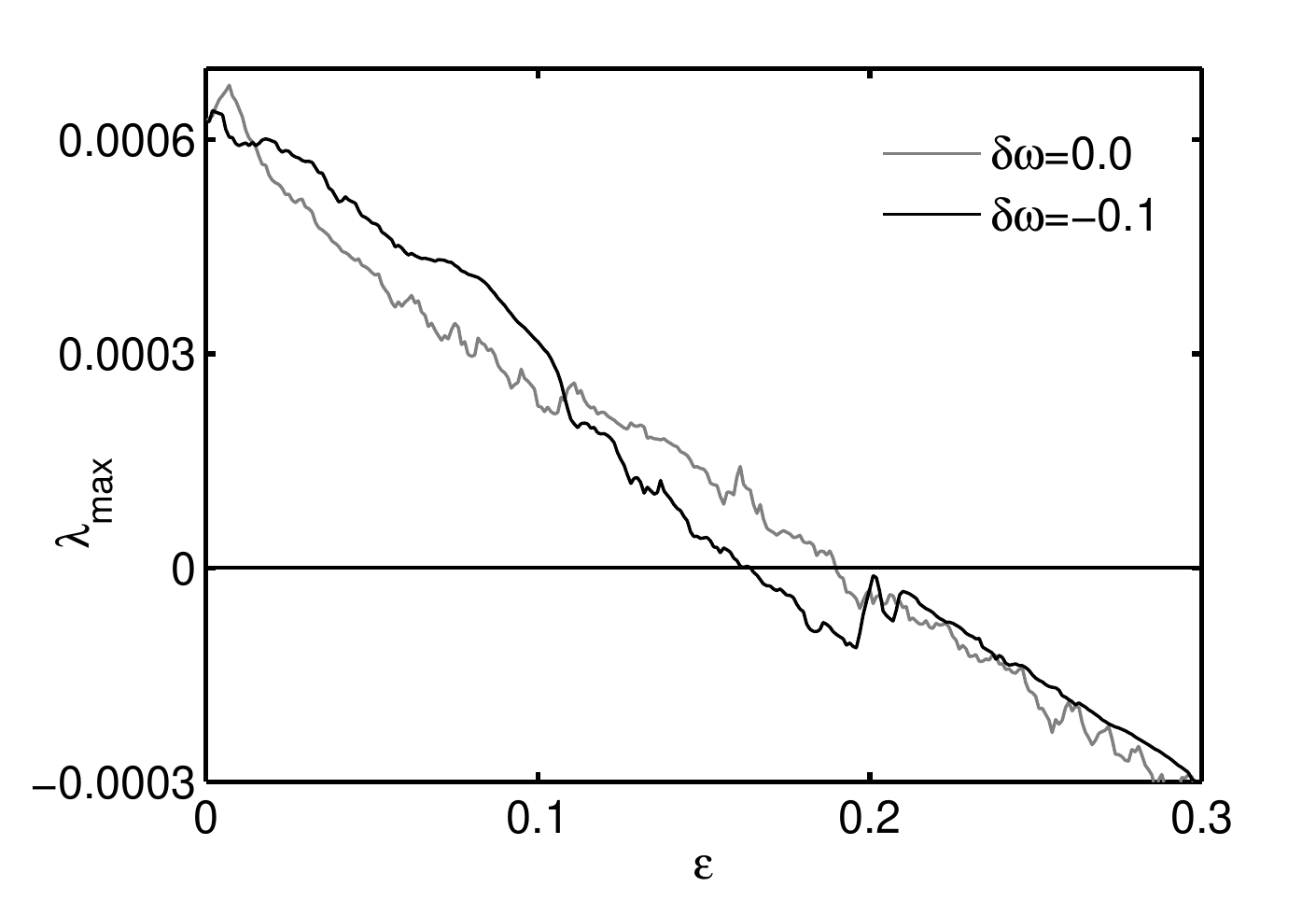} }
	\caption{ Plots of master stability function $\lambda_{max}$ against the coupling strength $\epsilon$ show (gray line for $\delta \omega$=0, black line for $\delta \omega=-0.1$) a drifting to a lower critical value for induced heterogeneity. For identical case, when $\delta \omega$=0 for all the oscillators in the lattice, $\omega_i$=1.1 is used and in the mismatched case, $\omega_i=1.1+\delta \omega$ is used only for the oscillator in black color. Rest of the oscillators have the same set of parameter values.}
	\label{neg-mis}
\end{figure}

\begin{figure}
	\resizebox{0.4\columnwidth}{!}{%
		\includegraphics[height=19cm,width=21cm]{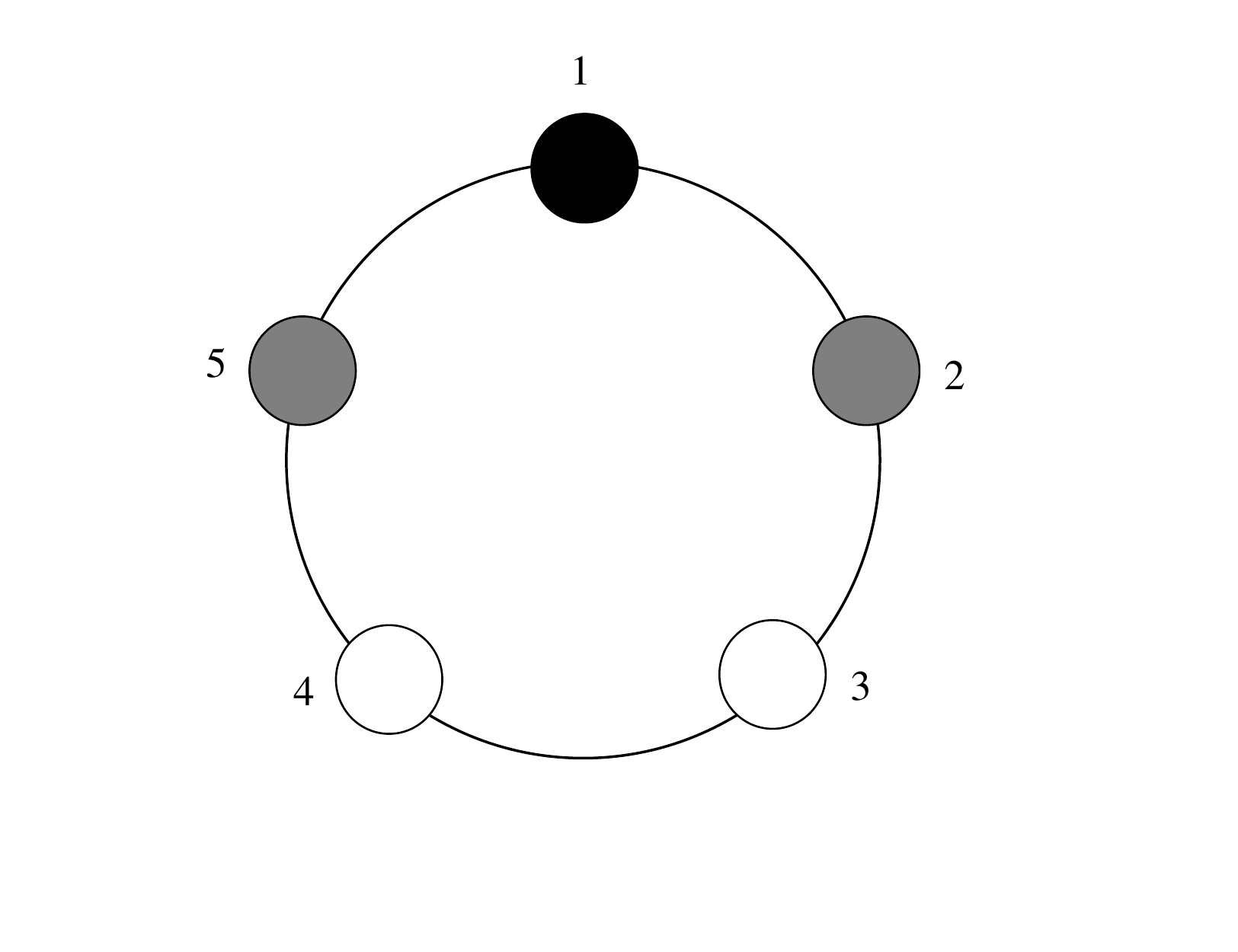} }
	{\includegraphics[height=4.5cm,width=6.5cm]{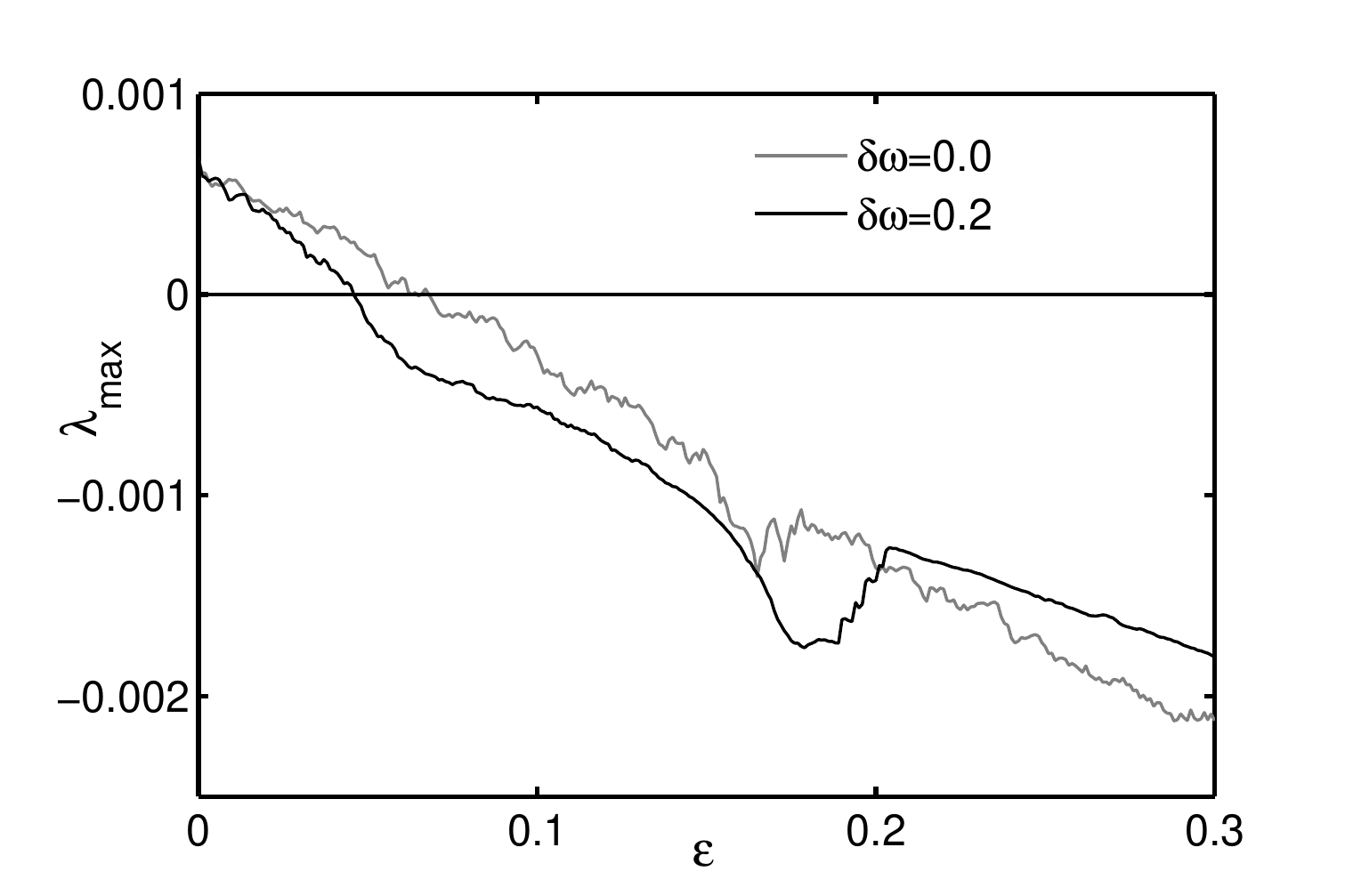} }
	\caption{Left panel shows five oscillators are in the form of a closed ring. The oscillator in black color has parameter mismatch. Right panel:  Variation of MSF with respect to coupling strength $\epsilon$.}
	\label{ring}     
\end{figure}
\begin{figure}
	\resizebox{0.4\columnwidth}{!}{%
		\includegraphics[height=20.0cm,width=20.0cm]{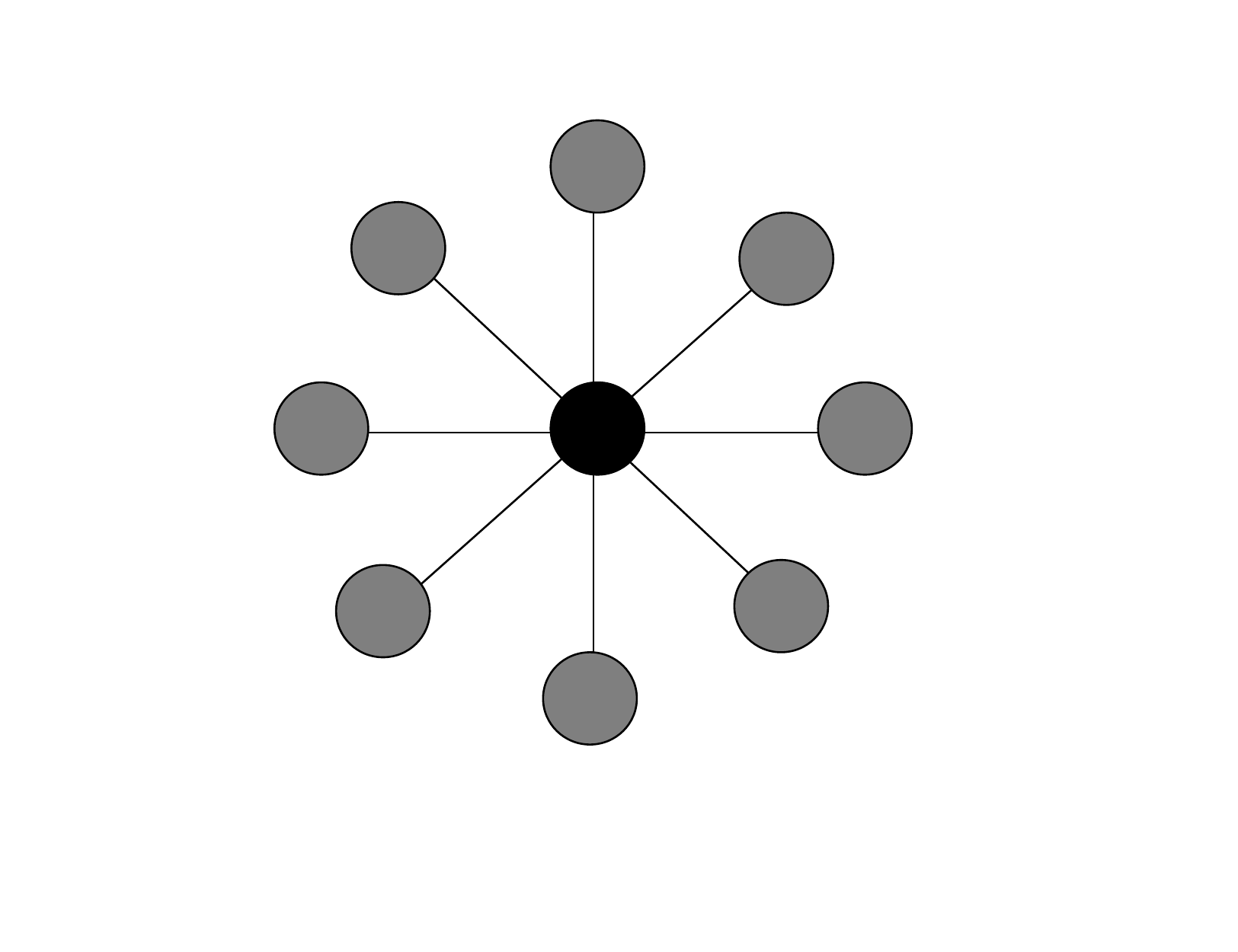} }
	{\includegraphics[height=5cm,width=7cm]{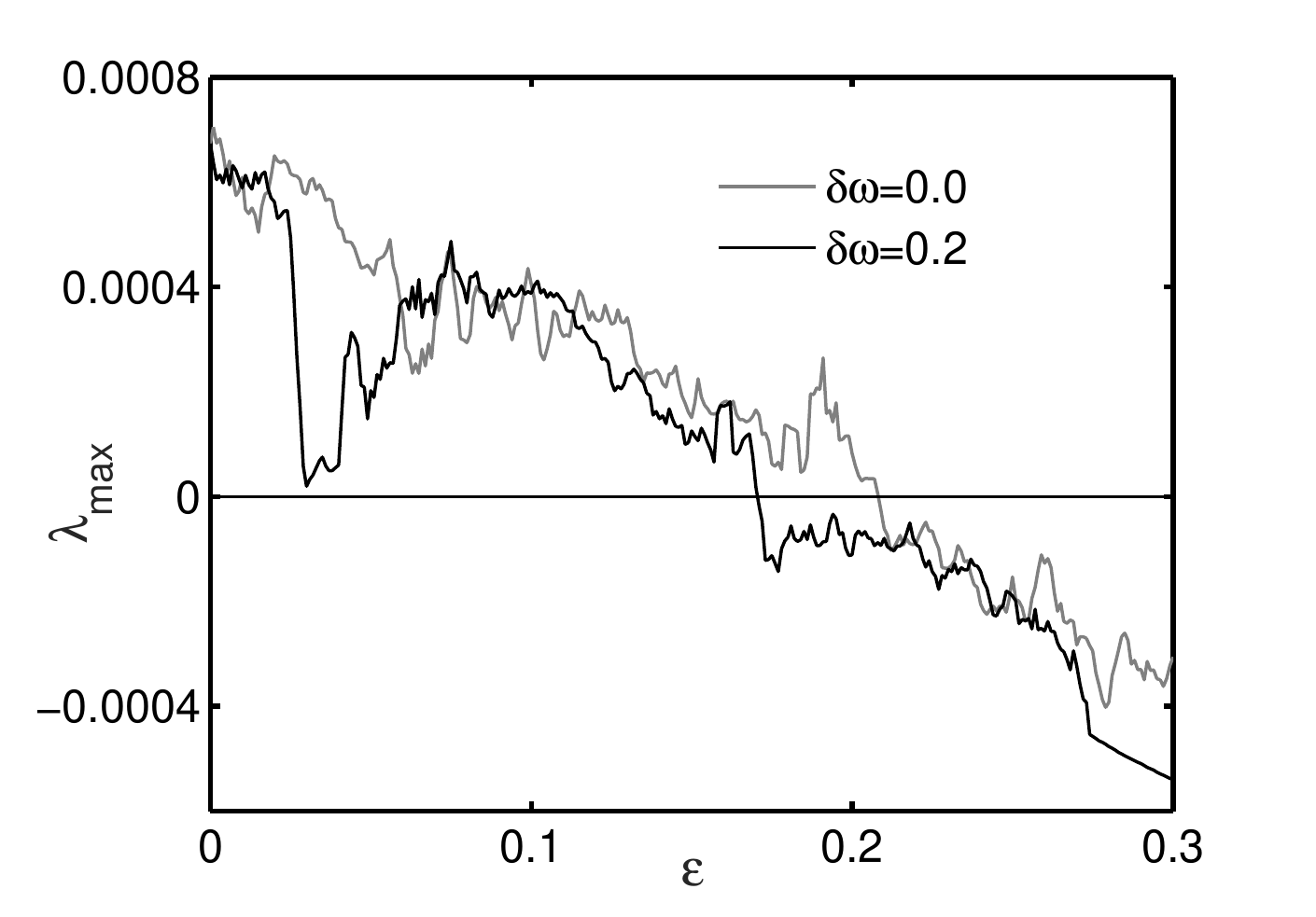} }
		\caption{Left panel: Coupling configuration of star network. All the gray color oscillators are identical and black oscillator has parameter mismatch. Right panel: Variation of MSF for synchronization between identical nodes (gray circle). }
	\label{star}     
\end{figure}

\section{Star Lattice}
\label{sec:2}
A similar enhancing of synchronization due to heterogeneity is also observed in a star network.  In a  star network shown in the left panel of Fig.~\ref{star}, one central node (black circle)  communicates directly with all the remaining nodes (gray color circles), and the other nodes communicate between themselves via the central node considered as a hub. The central node relay the information to all the other oscillators.  The heterogeneity is induced in the hub  as a mismatch $\delta\omega$. The dynamical equation of the star network is
$$\dot x_1=-\omega_1y_1-z_1+\frac{\epsilon}{N-1}\sum_{j=2}^{N}(x_{j}-x_1)\;\;\;\;\;\;\;\;\;\;\;\;\;\;\;\;\;\;\;\;\;\;\;\;\;\;\;\;\;\;\;\;$$
$$ \dot y_1=\omega_1 x_1+ay_1 \;\;\;\;\;\;\;\;\;\;\;\;\;\;\;\;\;\;\;\;\;\;\;\;\;\;\;\;\;\;\;\;\;\;\;\;\;\;\;\;\;\;\;\;\;\;\;\;\;\;\;\;\;\;\;\;\;\;\;\;\;\;\;\;\;\;\;\; \eqno{(2a)}$$
$$ \dot z_1 =b+z_1(x_1-c)\;\;\;\;\;\;\;\;\;\;\;\;\;\;\;\;\;\;\;\;\;\;\;\;\;\;\;\;\;\;\;\;\;\;\;\;\;\;\;\;\;\;\;\;\;\;\;\;\;\;\;\;~~~~~~~~~$$
$$\dot x_i=-\omega_iy_i-z_i+\epsilon(x_1-x_i)~~~~~~~~~~~~~~~~~~~~~~~~~~~~~~~~~~~~~~~~$$
$$ \dot y_i=\omega_i x_i+ay_i \;\;\;\;~~~~~~~~~~~~~~~~~~~~~~~~~~~~~~~~~~~~~~~~~~~~~~~~~~~~~~ \eqno{(2b)}$$
$$ \dot z_i =b+z_i(x_i-c),\;\;\;\;i=2,...,N.~~~~~~~~~~~~~~~~~~~~~~~~~~~~~~~~~~$$
Equation~(2a) describing the dynamics of $[x_1, y_1, z_1]^T$ represents the hub. Ae parameter mismatch  $\omega_1=1+\delta\omega$ is introduced in the hub. All indirectly connected outer nodes are identical ($\omega_i=1$ for $i=2,...,N$), which are represented by their state variables $[x_i, y_i, z_i]^T, i=2,3,...N$ in Eqn.(2b). To reveal enhancing of synchrony in the star network, we consider $N=15$ R\"ossler oscillators and their system parameters in a chaotic regime. Variation of $\lambda_{max}$ with coupling strength is shown in the right panel of Fig.~\ref{star} for identical (gray) and mismatched (black) cases. Figure~4(b) of $\lambda_{max}$ clearly shows a lowering of critical coupling (black) of CS in the outer oscillators for a mismatch in the hub. Plot of $\lambda_{max}$ for all identical case shows a larger critical coupling (gray) for CS.
\begin{figure}
	\resizebox{0.6\columnwidth}{!}{%
		\includegraphics{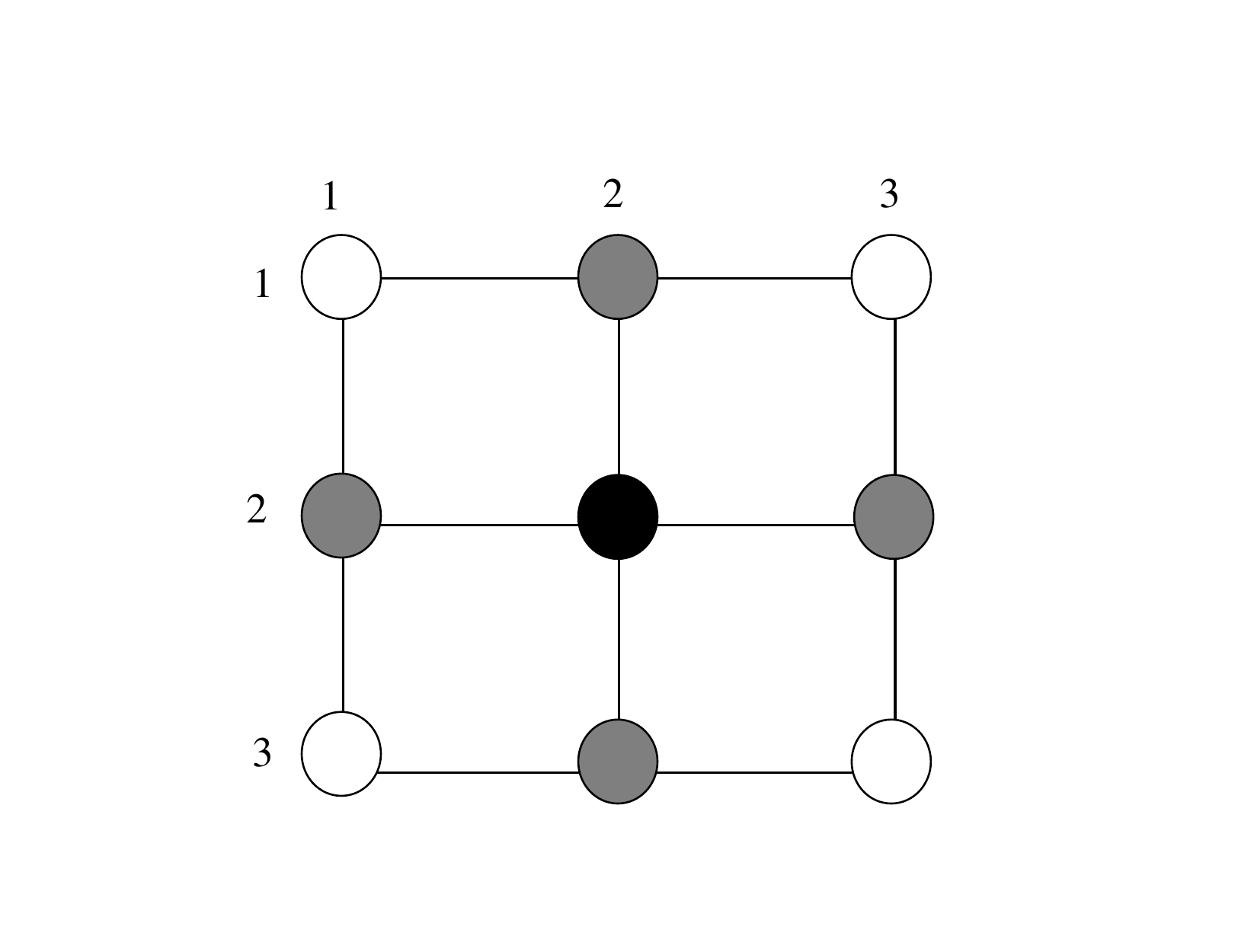} }
	\caption{Two dimensional grid of bidirectionally coupled R\"ossler oscillator where black color oscillator represents the mismatched oscillator and open circles and gray colors circles are identical oscillators.}
	\label{3x3_config}     
\end{figure}

\begin{figure}
	\resizebox{1.0\columnwidth}{!}{%
		\includegraphics{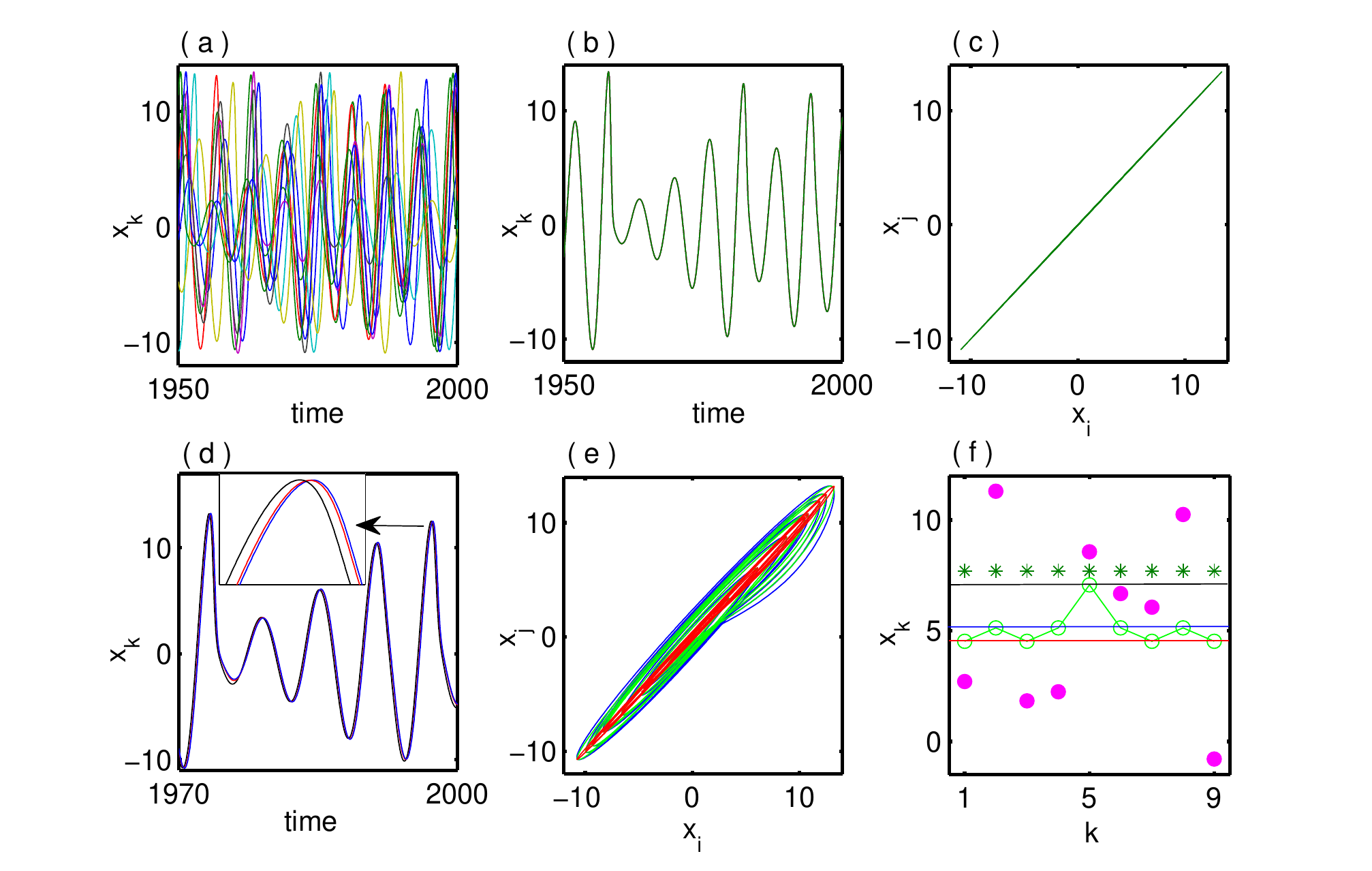} }
	\caption{Clustering synchronization in a 2D lattice R\"ossler oscillators. Time series of 9-identical oscillators show a desynchronized state for $\epsilon=0.19.$ in (a). In absence of parameter mismatch ($\delta \omega=0$), time series in (b) of all the oscillators are in CS for a larger $\epsilon=0.25$, corresponding synchronization manifolds in (c). Under induced heterogeneity $\delta\omega=0.1$ and the lower coupling strength, $\epsilon=0.19$, three clusters form. Two clusters emerge into separate CS states but they develop LS (blue and red lines) as shown in their time series plot in (d) when the central oscillator (black line) is isolated but having a lower lag time with the nearest neighbor. The corresponding LS manifolds are shown in (e). Snapshots of all the oscillators at a particular instant are plotted in (f) where magenta ( solid circles ), dark green (star) and green ( open circles ) colors represent the desynchronized state ($\delta\omega=0.0, \epsilon=0.19$), CS state ($\delta\omega=0.0, \epsilon=0.25$) and cluster synchronization ($\delta\omega=0.1, \epsilon=0.19$), respectively. In cluster state, two clusters green color (open circles) are seen along the blue and red lines, while the  isolated central node is seen along the black line.}
	\label{fig_two_D}     
\end{figure}
\begin{figure}
	\resizebox{0.5\columnwidth}{!}
	{\includegraphics{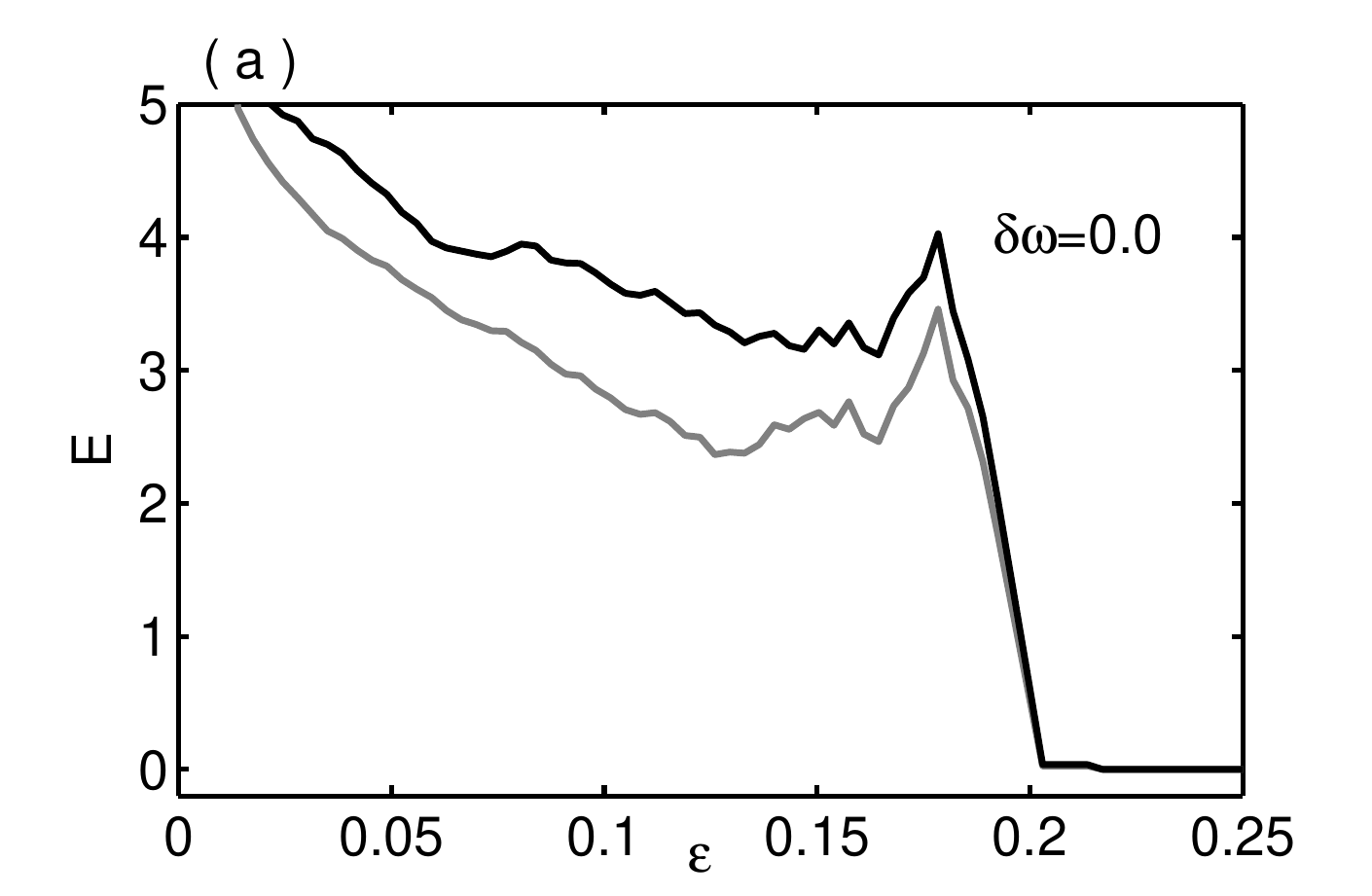} }\\
	\resizebox{0.5\columnwidth}{!}
	{\includegraphics{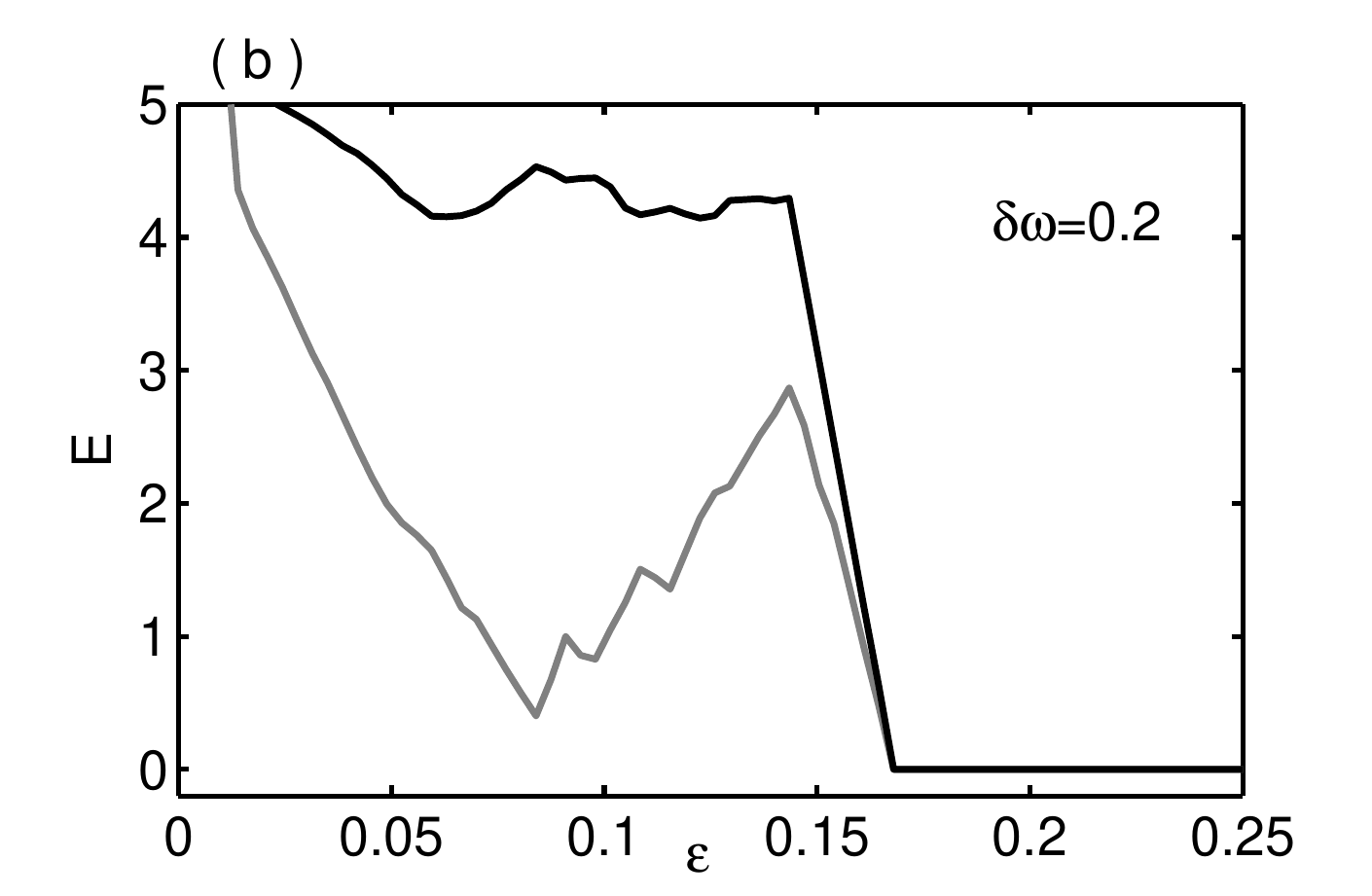} }\\
	\resizebox{0.5\columnwidth}{!}
	{\includegraphics{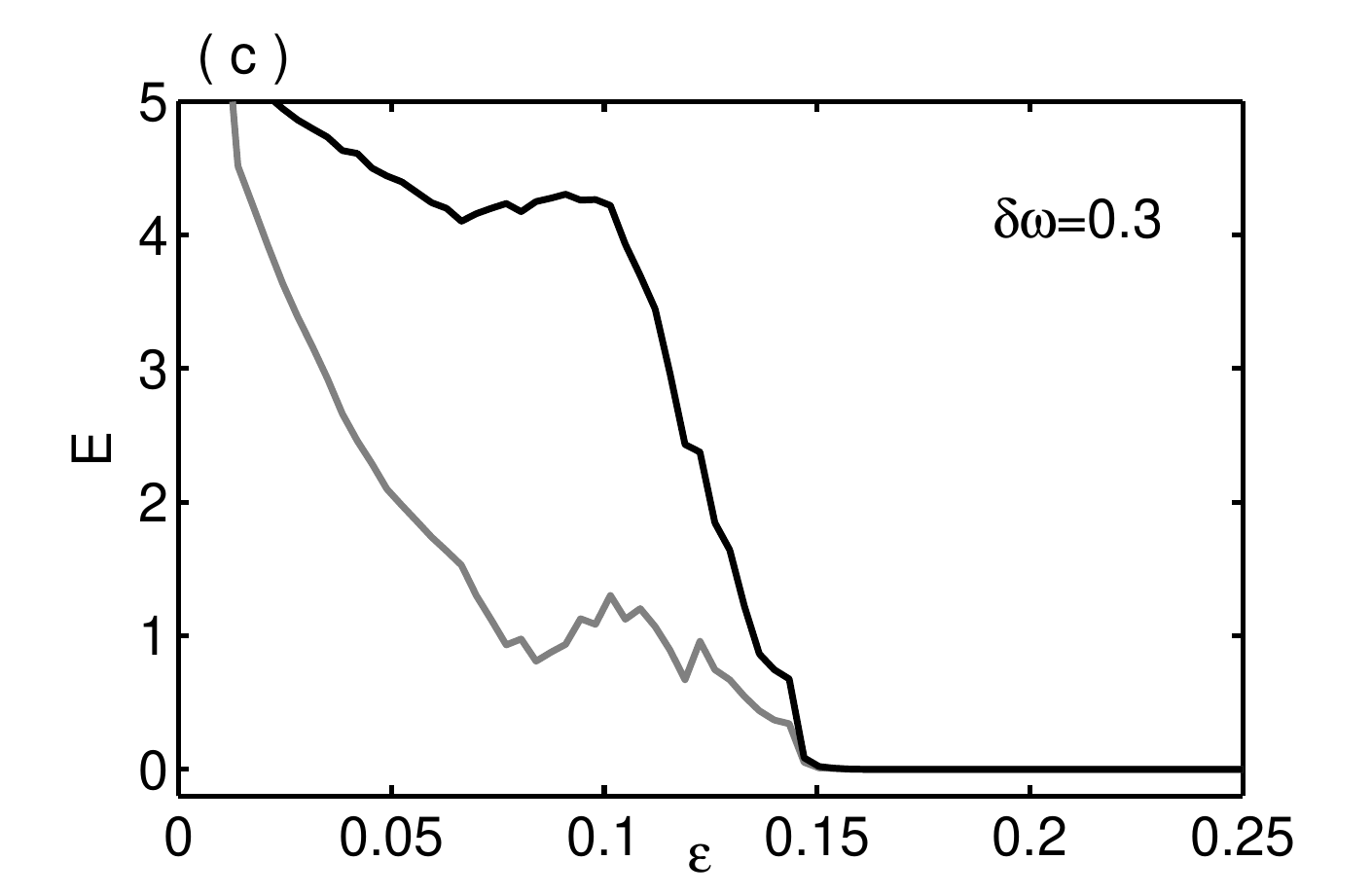} }\\
	\resizebox{0.5\columnwidth}{!}
	{\includegraphics{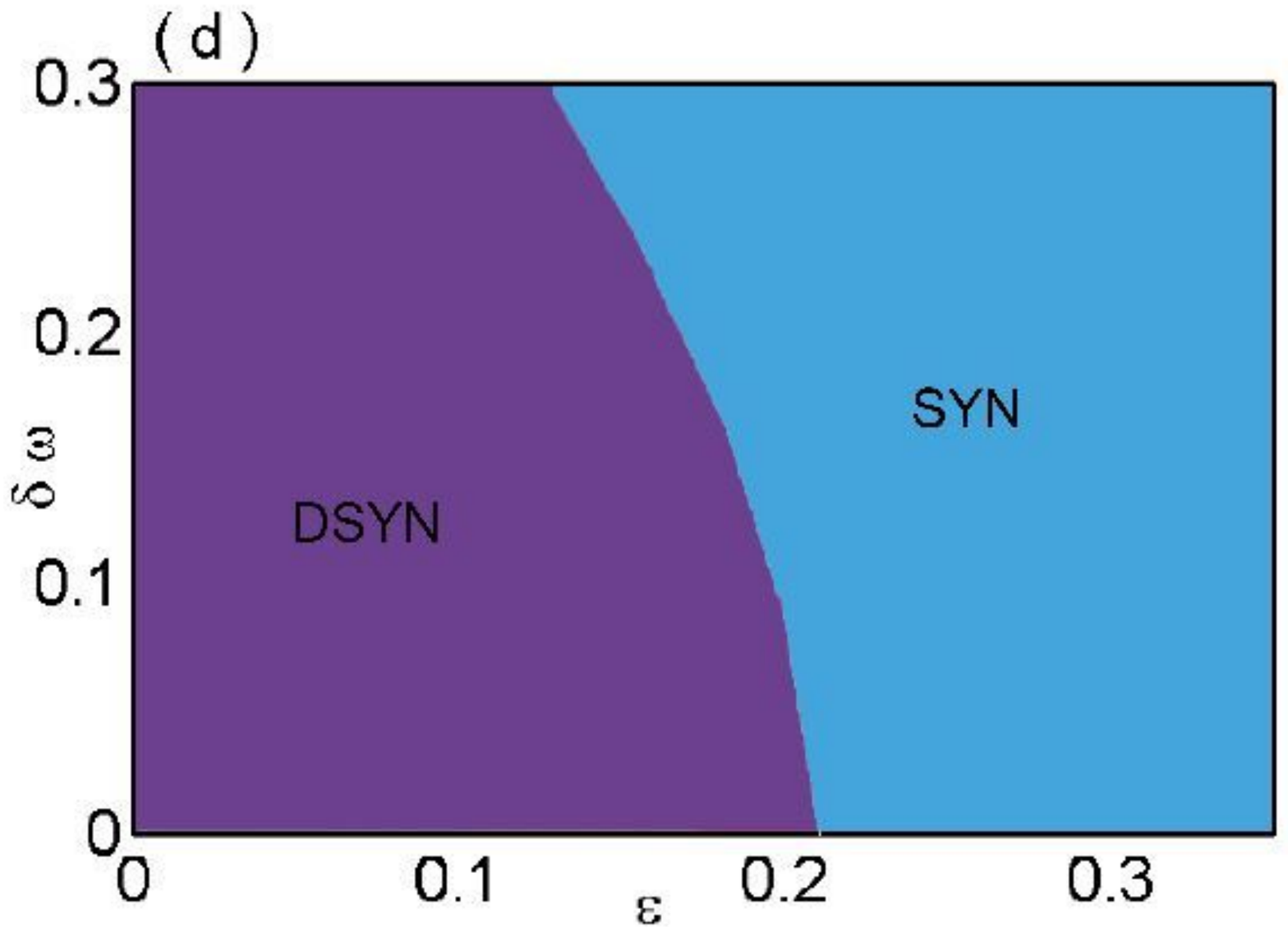} }
	\caption{(Color Online)  Synchronization error $E$ against coupling strength $\epsilon$ of the bidirectionally coupled 2D lattice,  (a) $\delta\omega=0.0$, (b) $\delta\omega=0.2$ and (c) $\delta\omega=0.3$. The black and gray colors represent the corresponding synchronization error of two cluster synchronization (open circle and gray color oscillators in 2D coupling topology). (d)  The enhancement of relay synchronization is shown by varying the coupling strength $\epsilon$ and mismatched parameter $\delta\omega$ in two dimensional phase plane $\epsilon - \delta \omega$ where violet (black) and cyan (gray) colors represent the corresponding desynchronized (DSYN) and synchronized (SYN) regions respectively.}
	\label{2d_grid}     
\end{figure}
\section{Two Dimensional Lattice}
Finally, we consider  a 2D lattice of chaotic R\"ossler oscillators shown in Fig.~\ref{3x3_config}. The dynamical equation of the 2D lattice is
$$\dot x_{i,j}=-\omega_{i,j}y_{i,j}-z_{i,j}+\epsilon(x_{i-1,j}+x_{i+1,j}+x_{i,j-1}+x_{i,j+1}-4x_{i,j})$$
$$ \dot y_{i,j}=\omega_{i,j}x_{i,j}+ay_{i,j}\;\;\;\;\;\;\;\;\;\;\;\;\;\;\;\;\;\;\;\;\;\;\;\;\;\;\;\;\;\;\;\;\;\;\;\;\;\;\;\;\;\;\;\;\;\;\;\;\;\;\;\;\;\;\;\;\;\;\;\;\;\;\;\;\;\;\;\;\;\;\; \eqno{(3)}$$
$$\dot z_{i,j}=b+z_{i,j}(x_{i,j}-c) \;\;\;\;\;\;\;\;\;\;\;\;\;\;\;\;\;\;\;\;\;\;\;\;\;\;\;\;\;\;\;\;\;\;\;\;\;\;\;\;\;\;\;\;\;\;\;\;\;\;\;\;\;\;\;\;\;\;\;\;\;\;\;\;\;\;\;\;$$ 
where $i,j=1, 2,3$. Individual oscillators are chaotic in absence of coupling  ($\epsilon=0.0$) for $\omega_{i,j}=1$ for all $i,j$ and parameters, $a=0.2, b=0.4, c=7.5$. As an example, we consider a $3\times3$ lattice in Fig.~\ref{3x3_config}: all open circles and gray circles represents  identical oscillators, $\omega_{i,j}=1$ and the central black node  represents the mismatched oscillator $\omega _{2,2}+\delta\omega$. All open circle nodes communicates with the black node via two links whereas the gray colors nodes form  direct links with the black node. For the identical case (i.e. $\delta \omega=0.0$), we consider a lower  coupling strength $\epsilon=0.19$ than the critical coupling when they are all desynchronized: chaotic time series of all identical oscillators confirm this in Fig.~\ref{fig_two_D}(a). When the coupling strength is increased to $\epsilon=0.25$ (above the critical value), all the oscillators reaches a CS state in chaotic motion as confirmed by their corresponding time series in Fig.~\ref{fig_two_D}(b) and the associated synchronization manifold $x_i$ vs $x_j$  in Fig.~\ref{fig_two_D}(c). When we  induce a mismatch in the hub (black circle) at a lower coupling $\epsilon=0.19$, we observe three separate groups or clusters: open circle nodes form a cluster in a CS state, gray color nodes form another cluster in a CS state but maintain a LS with the open circle nodes. On the other hand, the central hub remain isolated and forms a separate group, however, it also maintains a lag with gray color nodes but different from that of open circle nodes. We refer to this overall state as cluster synchronization.  For demonstration, we induce a small mismatch $\delta\omega=0.1$ in the central oscillator (black circle) when all $9$-oscillators are divided into three clusters: one cluster contains  degree  $2$ nodes (open circles) faraway from the central node (black circle), another cluster contains the degree $3$ nodes (gray color circles) and at a relay distance $1$ from the central node (black circle) and, the central mismatched oscillator remains isolated from other two clusters. The time series of all oscillators are plotted in Fig.~\ref{fig_two_D}(d) for $\delta\omega=0.1$ and $\epsilon=0.19$ which are magnified in  the inset where three distinct colors, namely, black, blue and red lines are shown. It indicates that all open circle nodes are in a CS state and, the gray color nodes are also in a separate CS state but they maintain a lag. The black line represents the hub node which also maintain LS with the other two clusters, however, its lag time is different from red and blue lines. 
The amount of lag time between the central node and the open circle nodes is more than the lag between the central node and the gray color nodes. The synchronization manifolds $x_{1,1}$ vs $x_{2,2}$ (blue color), $x_{1,2}$ vs $x_{2,2}$ (green color) and $x_{1,1}$ vs $x_{1,2}$ (red color) are plotted in Fig.~\ref{fig_two_D}(e). Snapshots of the amplitude at a particular instant are plotted in Fig.~\ref{fig_two_D}(f) where magenta ( solid circles ), green ( open circles ) and dark green ( stars ) colors represents the desynchronization ($\epsilon=0.19, \delta \omega=0.0$), cluster synchronization ($\epsilon=0.19, \delta \omega=0.1$) and CS ($\epsilon=0.25, \delta \omega=0.0$) in the network. 

\par Enhancing of synchronization between the indirectly coupled oscillators of the 2D lattice is described in Fig.~\ref{2d_grid}. In absence of any mismatch ($\delta \omega=0.0$), the two groups (open and gray color circles) are completely synchronized for $\epsilon \ge 0.22$. Figure~\ref{2d_grid}(a) shows the variation of synchronization errors $E$ of the open and gray color circles nodes by changing the coupling strength $\epsilon$. From this figure it is shown that the oscillators in the two groups are synchronized at the same critical coupling strength.  If we introduce heterogeneity in the central oscillator $(\delta\omega=0.2)$, the whole network forms two clusters (except the central node) and the oscillators (open and gray color circles) in each cluster  are fully synchronized for lower value of  $\epsilon \ge 0.17$. The corresponding synchronization error with respect to coupling strength is shown in Fig.~\ref{2d_grid}(b). By further increasing the heterogeneity at the relay unit by $\delta\omega=0.3$, synchronization in the two clusters is more enhanced (Fig.~\ref{2d_grid}(c)). To explore the whole scenario  by changing the heterogeneity $\delta \omega$ and the coupling strength $\epsilon$, we plot a phase diagram in $ \epsilon - \delta \omega $ plane in Fig.~\ref{2d_grid}(d) where violet (black) and sky (gray) color represent the desynchronized and synchronized region. The borderline of the synchronous and asynchronous regions indicates the critical coupling strength. It is clear that the critical coupling $\epsilon=\epsilon_c$ decreases with increase of mismatch $\delta \omega$.

\section{Conclusion}
The effect of heterogeneity was studied on CS of chaotic oscillators using different configurations ranging from 1D array, ring, star and 2D lattice. It was shown that an induced heterogeneity or parameter mismatch in a suitably located central oscillator works as a relaying device to enhance CS in other identical nodes which interacted via the central node. The effect was found general in all the given example networks. The heterogeneity in the form of a parameter mismatch induced a LS between the mismatched oscillator and all the other identical oscillators, which, in turn, helped the identical oscillators at symmetric positions with respect to the relay unit, synchronize at a lower coupling strength than the critical coupling of CS in all identical oscillators. The time-lag remained unaffected by the distance of the relayed units from the relay unit in 1D open array. In case of 2D lattice the time-lag changed with the degree of the relayed units and thereby they evolved into cluster synchronization. Each cluster remained in CS states but with separate lag between different clusters.\\
\\
\\
\par {\bf Acknowledgments:}
D.G. was supported by SERB-DST (Department of Science and Technology), Government of India (Project no. EMR/2016/001039). S.K.D. also acknowledges support by the Emeritus Fellowship of the University Grants Commission (India).


\begin{thebibliography}{111}
\bibitem{bocaletti} S. Boccaletti, {\it The Synchronized Dynamics of Complex Systems,} \textbf{6}, Elsevier, Amsterdam, (2008)
\bibitem{application} L.M. Pecora, T.L. Carroll, G.A. Johnson, D.J. Mar, J.F. Heagy, Chaos \textbf{7(4)}, 520 (1997)
\bibitem{communi1} K. Murali, H. Yu, V. Varadan H. Leung, IEEE Transactions on Consumer Electronics \textbf{47}, 709 (2001)
\bibitem{communi2} A.N. Miliou, I.P. Antoniades, S.G. Stavrinides, A.N. Anagnostopoulos, Nonlinear Analysis: Real World Applications \textbf{8}, 1003 (2007)
\bibitem{communi_pre} D. Ghosh, S. Banerjee, Phys. Rev. E \textbf{78}, 056211 (2008)
\bibitem{communi_epl1} D. Ghosh, S. Banerjee, A.R. Chowdhury, Euro. Phys. Lett. \textbf{80}, 30006 (2007)
\bibitem{communi_epl2} S. Banerjee, D. Ghosh, A. Ray, A.R. Chowdhury, Euro. Phys. Lett. \textbf{81}, 20006 (2008)		
\bibitem{kurths} A. Pikovsky, M. Rosemblum, J. Kurths, {\it Synchronization: A Universal Concept in Nonlinear Sciences,} \textbf{12}, (Cambridge University
Press, Cambridge, 2003)
\bibitem{phys_report} S. Boccaletti, J. Kurths, G. Osipov, D. L. Valladares, C. S. Zhou, phys. Rep. \textbf{366}, 1 (2002)
\bibitem{relay1} I. Fischer, R. Vicente, J.M. Buld\'{u}, M. Peil, C.R. Mirasso, M.C. Torrent, J. Garc\'{i}a-Ojalvo, Phys. Rev. Lett. \textbf{97}, 123902 (2006)
\bibitem{phase_prl} M.G. Rosenblum, A.S. Pikovsky, J. Kurths, Phys. Rev. Lett. \textbf{76}, 1804 (1996)
\bibitem{lag_prl} M.G. Rosenblum, A.S. Pikovsky, J. Kurths, Phys. Rev. Lett. \textbf{78}, 4193 (1997)
\bibitem{genl_pre} N.F. Rulkov, M.M. Sushchik, L.S. Tsimring, H.D.I. Abarbanel, Phys. Rev. E \textbf{51}, 980 (1995)
\bibitem{cluster} V.N. Belykh, I.V. Belykh, E. Mosekilde, Phys. Rev. E \textbf{63}, 036216 (2001)
\bibitem{partial} C.V. Vreeswijk, Phys. Rev. E \textbf{54}, 5522 (1996)
\bibitem{chimera} D.M. Abrams S.H. Strogatz, Phys. Rev. Lett. \textbf{93}, 174102 (2004)
\bibitem{chimera2} B.K. Bera, D. Ghosh, M. Lakshmanan, Phys. Rev. E \textbf{93}, 012205 (2016)
\bibitem{chimera3} B.K. Bera, D. Ghosh, Phys. Rev. E \textbf{93}, 052223 (2016)
\bibitem{relay2} R. Banerjee, D. Ghosh, E. Padmanaban, R. Ramaswamy, L. M. Pecora, S. K. Dana, Phys. Rev. E  \textbf{85}, 027201 (2012)  
\bibitem{relay3} A. Bergner, M. Frasca, G. Sciuto, A. Buscarino, E.J. Ngamga, L. Fortuna, J. Kurths, Phys. Rev. E \textbf{85}, 026208 (2012)
\bibitem{relay4} R. Guti\'{e}rrez, R. Sevilla-Escoboza, P. Piedrahita, C. Finke, U. Feudel, J. M. Buld\'{u}, G. Huerta-Cuellar, R. Jaimes-Reategui,
Y. Moreno, S. Boccaletti, Phys. Rev. E \textbf{88}, 052908 (2013)
\bibitem{remote1} A. Bergner, M. Frasca, G. Sciuto, A. Buscarino, E.J. Ngamga, L. Fortuna, J. Kurths, Phys. Rev. E \textbf{85}, 026208 (20012)
\bibitem{remote2} L. V. Gambuzza, A. Cardillo, A. Fiasconaro, L. Fortuna, J. G\'{o}mez-Gardenes, M. Frasca, Chaos \textbf{23}, 043103 (2013)
\bibitem{laser1} A. Englert, W. Kinzel, Y. Aviad, M. Butkovski, I. Reidler, M. Zigzag, I. Kanter, M. Rosenbluh, Phys. Rev. Lett. \textbf{104}, 114102 (2010)	
\bibitem{laser2} J. Tiana-Alsina, K. Hicke, X. Porte, M.C. Soriano, M.C. Torrent, J. Garcia-Ojalvo, I. Fischer, Phys. Rev. E  \textbf{85}, 026209 (2012) 
\bibitem{laser3} J.G. Wu, Z.M. Wu, G.Q. Xia, T. Deng, X.D. Lin, X. Tang, G.Y. Feng, IEEE Phot. Tech. Lett. \textbf{23}, 1854 (2011)
\bibitem{electronic1} I.G. Da Silva, J.M. Buld\'{u}, C.R. Mirasso, J. Garc\'{i}a-Ojalvo, chaos  \textbf{16}, 043113 (2006)
\bibitem{electronic2} A. Wagemakers, J.M. Buld\'{u}, M.A.F. Sanju\'{a}n, chaos  \textbf{17}, 023128 (2007)
\bibitem{communi3} A. Wagemakers, J.M. Buld\'{u}, M.A.F. Sanju\'{a}n, Europhys. Lett. \textbf{81}, 40005 (2008)
\bibitem{brain1} A.K. Engel, P. K\"{o}nig, A.K. Kreiter, W. Singer, Science  \textbf{252}, 1177 (1991)
\bibitem{brain2} P. R. Roelfsema, A.K. Engel, P. K\"{o}nig, W. Singer, Nature \textbf{385}, 157 (1997)
\bibitem{brain3} R. Vicente, L.L. Gollo, C.R. Mirasso, I. Fischer, G. Pipa, Proc. Natl. Acad. Sci. USA \textbf{105}, 17157 (2008)
\bibitem{enh_boca}L.V. Gambuzza, M. Frasca, L. Fortuna S. Boccaletti, Phys. Rev. E  \textbf{93}, 042203 (2016) 
\bibitem{enh_delay} M. Dhamala, V. K. Jirsa, and M. Ding, Phys. Rev. Lett. \textbf{92}, 074104 (2004)
\bibitem{diba_physica} A. Ray, A. Roy Chowdhury, D. Ghosh, Physica A \textbf{392}, 4837 (2013)
\bibitem{diba} S. Majhi, B.K. Bera, S. Banerjee, D. Ghosh, Eur. Phys. J. Special Topics \textbf{225}, 65 (2016)
\bibitem{mixed_syn} S.K. Bhowmick, B.K. Bera, D. Ghosh, Commun. Nonlinear Sci. Numer. Simulat. \textbf{22}, 692 (2015) 
\bibitem{senthil} D. V. Senthilkumar, J. Kurths, Eur. Phys. J. Special Topics \textbf{187}, 87 (2010)




\end{thebibliography}
\end{document}